\documentclass[twocolumn,showpacs]{revtex4}
\usepackage{amssymb}
\usepackage{amsmath}
\usepackage[dvips]{graphicx}

\begin{document}

\title{Extraordinary magnetoresistance in graphite: experimental evidence for the time-reversal symmetry breaking.}
\author{ Y.  Kopelevich, R. R. da Silva,  and A. S. Alexandrov$^{\star}$}
\affiliation{Instituto de Fisica "Gleb Wataghin", Universidade Estadual de Campinas, UNICAMP 13083-970, Campinas, Sao Paulo, Brasil}

\begin{abstract}
   We report an unusual highly-anisotropic  in-plane magneto-resistance (MR) in  graphite. In a certain current direction MR is negative and linear in low fields  with a crossover to a positive MR at higher fields, while in a perpendicular current direction we observe a giant super-linear  positive MR.  These extraordinary MRs are respectively explained by a hopping magneto-conductance via non-zero angular momentum orbitals, and by the  magneto-conductance of inhomogeneous media.  The linear negative orbital MR is a unique signature of the broken time-reversal symmetry (TRS) in graphite.  While some  local paramagnetic centers  could be  responsible for the broken TRS, the observed large diamagnetism suggests a more intriguing mechanism of  this breaking, involving    superconducting clusters  with  unconventional (chiral)  order parameters and spontaneously
generated normal-state current loops in graphite.
\end{abstract}


\maketitle

The transverse magneto-resistance (MR) in a magnetic field \textbf{B} along the z-axis   defined as $MR= \sigma_{xx}\sigma_{xx}(0)/(\sigma_{xx}^2+\sigma_{xy}^2)-1$ in terms of the conductivity tensor $\sigma_{ik}(B)$, is a sensitive probe of the electron transport mechanism in doped semiconductors and metals. In an isotropic medium with the Bloch electrons it is positive and  quadratic in B, except  the open Fermi surfaces,   where for
some specific directions of $\textbf{B}$ the positive MR is linear in B.  In the hopping regime with localized carriers MR is
caused by a strong magnetic field dependence of the exponential
asymptotic of  bound state wave functions at a remote distance
from a donor (or an acceptor),
which leads to a significant positive MR,  which is also  quadratic in low magnetic fields \cite{shklo}.

However, there is a substantial number of semiconductors and semimetals (e. g. bismuth) where  open Fermi surfaces are not feasible, but the positive MR is  linear. One of the theoretical possibilities for such a phenomenon is a so-called "quantum magnetoresistance" in semimetals  having  pockets of the Fermi surface with a small or even zero effective mass (as the Dirac fermions in graphite and graphene), which might be in the ultra-quantum limit at rather low magnetic fields \cite{abr}.

On the other hand,  there is anomalous \emph{negative} MR (NMR) first observed in some hopping systems, for
instance in amorphous germanium and silicon.  More recently a linear  NMR has been observed in the longitudinal c-axis interlayer current  in graphite at high magnetic fields
assigned  to  a growing population of the zero-energy Landau level of quasi-two-dimensional Dirac fermions with the increasing  magnetic field \cite{kop}. 

Here we report a low-field  transverse  MR in bulk quasi-two-dimensional graphite samples, which is giant ( 1400$\%$ at 2K and $B=200$  mT) and positive in one in-plane current direction, and linear and negative (below $80$ mT) in the other  direction, when the magnetic field is applied perpendicular to the planes. These observations are quantitatively explained in the framework of the magnetotransport of inhomogeneous low-carrier semiconductors \cite{dykhne} and of a hopping magneto-conductance via non-zero angular momentum orbitals \cite{aleomr}.

We have measured  10 commercially available HOPG samples with different mosaicity, characterized by FWHM of X-ray rocking curves, and the room temperature out-of-plane/basal plane zero-field resistivity ratio $\rho_c/\rho_{ab}$ ranging from $10^3$ to $10^5$. X-ray diffraction  measurements revealed a characteristic hexagonal graphite structure in the Bernal ($ABAB…$) stacking configuration, with no signature for other phases. The obtained crystal lattice parameters are $a = 2.48 \AA$  and $c = 6.71 \AA $. The results given in this letter were obtained on most anisotropic  ($\rho_c/\rho_b = 2 \times 10^5$)  HOPG samples from Union Carbide Co. and here shown for the sample with $ \rho_b = 5 \mu \Omega cm$,  $\rho_c = 1 \Omega cm$, and FWHM = 0.7$^o$ (the sample size is $l\times w \times t = 2.5 \times 2.5 \times 0.5 mm^3$). The magnetic field was applied parallel to the hexagonal c-axis ($B \parallel c \parallel t$), and  $\rho_{ab}(B,T)$ were recorded using the van der Pauw geometry (see insets in Fig. 1) sweeping the field between  a target negative and positive values using Quantum Design and Janis  9 T-magnet He-4 cryostats. For the measurements, silver past electrodes were placed on the sample surface, while the resistivity values were obtained in a geometry with a uniform current distribution through the sample cross section. All resistance measurements were performed in the Ohmic regime. Complementary magnetization measurements $M(B,T)$ were carried out by means of a SQUID (Quantum Design) magnetometer.

\begin{figure}
\begin{center}
\includegraphics[angle=-0,width=0.40\textwidth]{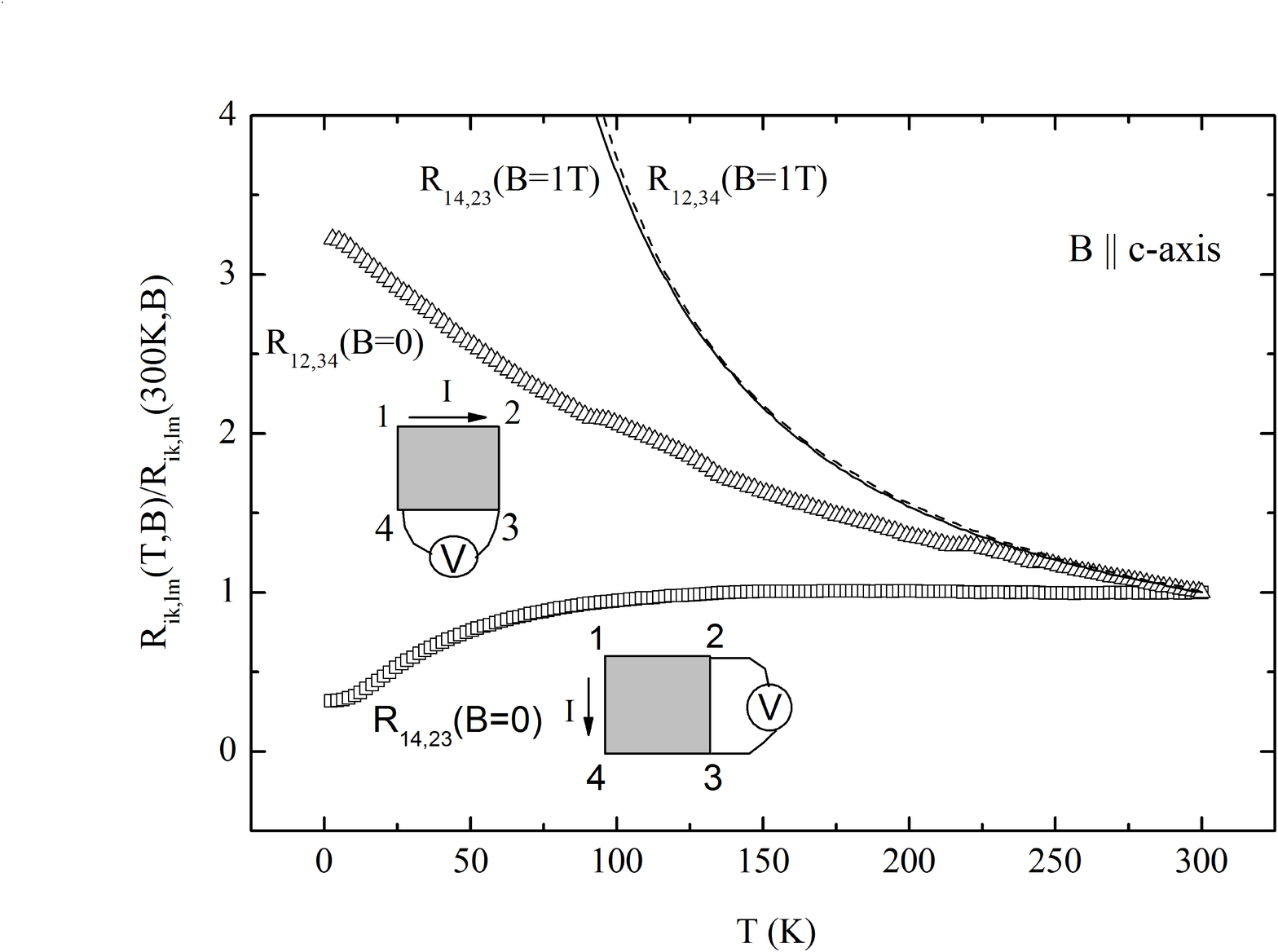}
\includegraphics[angle=-0,width=0.31\textwidth]{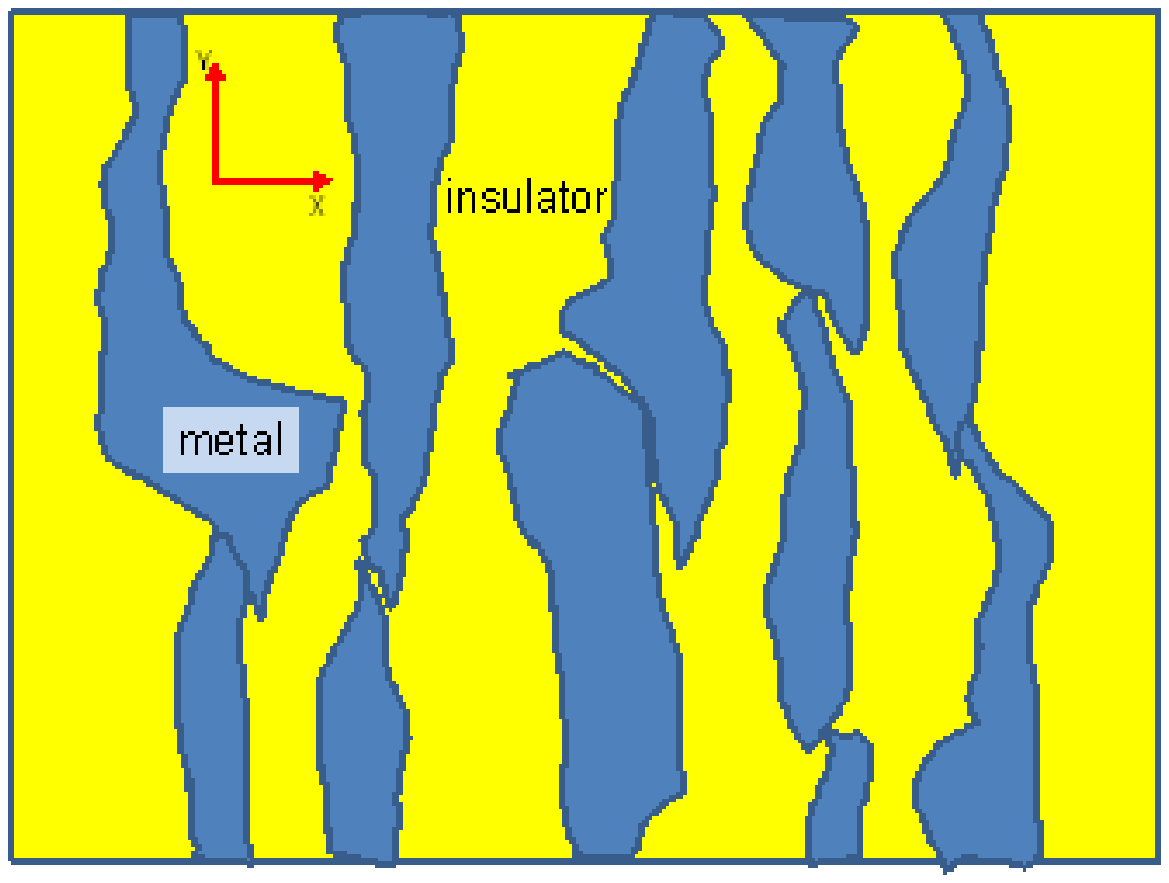}
\vskip -0.5mm \caption{(Color online) Upper panel: Insulating-like (X) and metallic-like (Y) zero-field in-plane resistance of  graphite  in two perpendicular current directions. The resistance becomes isotropic and insulating-like in high magnetic fields. Lower panel: well doped metallic clusters (dark regions) overlapped in Y-direction in an insulating matrix (light regions).}  \label{tempdep}
\end{center}
\end{figure}

The graphite samples described above show qualitatively different temperature dependence of the in-plane zero-field resistance for the current in two perpendicular in-plane directions,  Fig.(\ref{tempdep}). In one direction (called here  Y) the resistance is metallic-like
while in the other direction (X) it is insulating-like in a wide temperature interval above 2K. Since a sufficiently strong magnetic field makes graphene planes electrically isotropic, Fig.(\ref{tempdep}), the anisotropy is of electronic origin.   Most probably it is associated with the inhomogeneous carrier-density distribution, such that  metallic clusters are partially overlapped in Y-direction while they are separated by poorly doped insulating regions in X-direction, as illustrated in the lower panel of Fig.(\ref{tempdep}).  In high magnetic fields about 1 Tesla and higher resistance becomes isotropic and insulating-like in all directions, Fig.(\ref{tempdep}), because of a giant positive MR of the metallic clusters and suppression of possible  superconductivity, as discussed below.

\begin{figure}
\begin{center}
\includegraphics[angle=-0,width=0.40\textwidth]{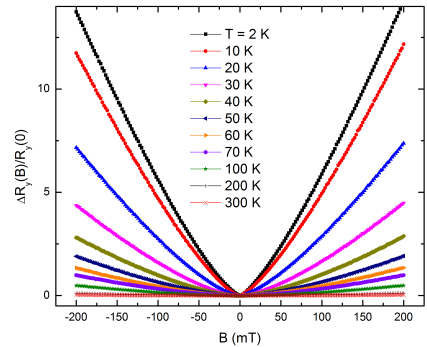}
\vskip -0.5mm \caption{(Color online)   Magnetoresistance in the metallic (Y) direction at different temperatures (symbols).}  \label{mry}
\end{center}
\end{figure}

When a relatively weak magnetic field is applied perpendicular to the planes, the MR along the metallic Y direction appears huge and positive, Fig.(\ref{mry}). A giant positive MR is naturally expected in doped graphite with  virtually massless Dirac fermions \cite{koplyk} since the parameter $\beta=\omega_c \tau$ becomes large already in the mT-region of the field (here $\omega_c$ and  $\tau$ are the Larmor frequency and the scattering time, respectively). However, it is neither quadratic as  in the Boltzmann theory, nor linear in B as in the "quantum magnetoresistance" \cite{abr}.  We suggest that inhomogeneities are responsible for the strong departure from these  regimes.
They lead to a radical rearrangement of the current flow pattern  changing the magnetic field dependence of the transverse conductivity in the strong magnetic field, $\beta > 1$ \cite{dykhne}. Importantly, when  $\sigma_{xy}\gg \sigma_{yy}$, then  even  relatively small inhomogeneities in the carrier density   lead to  the MR  proportional to $B^{4/3}$ \cite{dykhne}. In fact,  $MRy=(B/B_{in})^{4/3}$ fits nicely the observed magnetic field dependence of the magnetoresistance in the metallic current direction, Fig.(\ref{fit}) with a single scaling parameter $B_{in}$ depending on the fluctuations in the carrier density \cite{dykhne}.

\begin{figure}
\begin{center}
\includegraphics[angle=-0,width=0.38\textwidth]{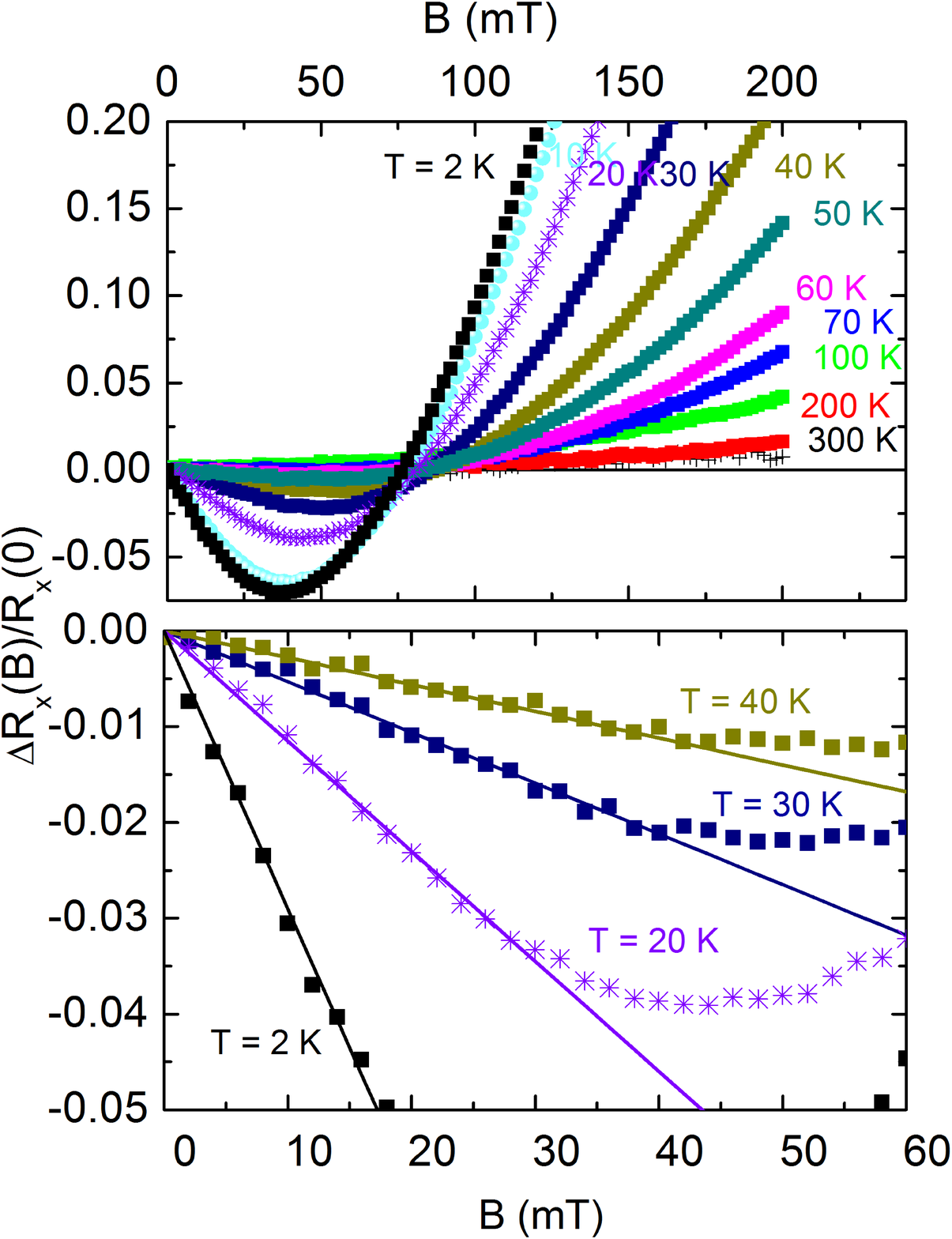}
\vskip -0.5mm \caption{(Color online)  Upper panel: Magnetoresistance in the insulating (X) direction at different temperatures (symbols). Lower panel shows the low-field   negative MR fitted  as $MRx= - c B$   with $c =
 0.0029$ 1/mT (T=2K), $c = 0.00115$ 1/mT (T=20K), $c = 0.00053$ 1/mT (T=30K), $c = 0.00028$  1/mT (T=40K),  indicating the time-reversal symmetry breaking.} \label{mrx}
\end{center}
\end{figure}

\begin{figure}
\begin{center}
\includegraphics[angle=-0,width=0.39\textwidth]{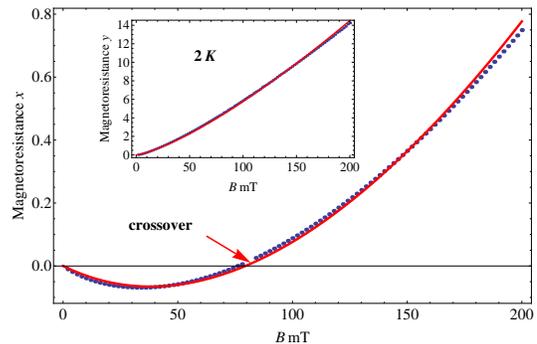}
\vskip -0.5mm \caption{(Color online)   Magnetoresistance in the insulating (X) direction at $T=2$K (symbols) fitted by Eq.(\ref{mrxtheory}) with $r=8.5$ and $B_{h}=110$ mT (line).  Inset: magnetoresistance in the metallic (Y) direction at $T=2$K (symbols) fitted by $(B/B_{in})^{4/3}$ with  $B_{in}=26.9$ mT (line).   } \label{fit}
\end{center}
\end{figure}

 The MR in the insulating direction is even more anomalous, Fig.(\ref{mrx}). It is  linear and negative in very low fields and positive and superlinear in higher fields above the crossover point. The insulating-like temperature dependence of the zero-field resistance, Fig.(\ref{tempdep}), in this direction supports the view that the metallic clusters are virtually nonoverlaped along X, so that the resistance is  the sum of the hopping resistance, $R_{h}$ of the insulating regions with low doping, and the metallic resistance $R_{m}$. Hence, the magnetoresistance in X direction is expressed via metallic  MRy and the hopping magnetoresistance, MRh of insulating layers as
 \begin{equation}
 MRx= {MRy\over{1+r}}+{MRh\over{1+1/r}}, \label{mrxtheory}
 \end{equation}
 where $r=R_h(0)/R_m(0)$ is the ratio of the zero-field insulating resistance to the zero-field metallic resistance. This ratio is about $r\approx 8.5$ at $2K$, Fig.(\ref{tempdep}).

In   inorganic and organic insulators  lattice defects such as
vacancies,  interstitials, excess  atoms or ions and other
"impurities" often localize carriers with a finite momentum rather
than in the zero-momentum s-states \cite{physics}. To model the hopping MRh  we apply a recent development of the conventional
theory of hopping  magnetoresistance for hopping via non-zero
orbital momentum states \cite{aleomr}. Quite remarkably this renders  a large weak-field hopping
MRh, which is negative. The theory \cite{aleomr} has accounted for the unconventional  MRs in some organic insulators (a so-called OMAR ) observed  in about the same range of the magnetic field \cite{omr} as in our graphite samples.  Moreover,  if the orbital
degeneracy is lifted due to a broken time-reversal symmetry
with or without net magnetization, the negative MRh is
\emph{linear} in $B$\cite{aleomr},
\begin{equation}
 MRh = \exp \left[-m \kappa r
b+{\kappa^3 \rho^2 r b^2\over{48}}\right]-1. \label{expansion}
 \end{equation}
Here $m$ is the magnetic quantum number of the localized state, $\kappa$ is the inverse zero-field localization length, $r$ and $\rho$ are an average hopping range and its projection on the plane perpendicular to the magnetic field, respectively, and  $b=B/B_0$
is the reduced magnetic field with $B_0=\hbar\kappa^2/2e$.

For the s-wave bound states with $m=0$ Eq.(\ref{expansion}) is
the textbook result \cite{shklo} predicting the positive hopping MRh. On the contrary,
for hopping via the orbitals with nonzero orbital momentum MRh
 is negative and linear in relatively small $B$, if $m=1,2,3,...$. The linear NMR is caused by the expansion of
the localized states with the positive $m$ due to a linear
magnetic lowering of their ionization energy by $\hbar \omega_c m/2$. It
 dominates in the wide region of realistic
impurity densities for any nonzero  $m$ because of the small numerical factor ($1/48$) in the quadratic term in the exponent  of Eq.(\ref{expansion}).

Using $MRh \approx \exp (-B/B_h)-1$ in Eq.(\ref{mrxtheory}) fits remarkably well the low-temperature MR in the insulating current direction with a single scaling parameter $B_h$  proportional to the localized-state ionization energy, Fig.(\ref{fit}). The same Eq.(\ref{mrxtheory})   describes reasonably well the  negative to positive MR crossover even at temperatures as high as $50$K. With increasing temperature deeper localized states with a higher ionization energy become accessible  for the hopping conductance, so that the slope  $c=1/B_h$ of the linear negative MR drops, as observed, Fig.(\ref{mrx}).

While the outlined model with one or two scaling parameters provides accurate agreement with our experimental observations, there are other theoretical mechanisms of NMR.
Importantly, there is no NMR and virtually no MR in the same range of the magnetic field \emph{parallel} to the planes, which rules out a spin origin of the observed MRs.
 Also there are no quantum magnetic oscillations at  high temperatures, where  unusual MRs are still observed, Figs.(\ref{mry},\ref{mrx}), so they are not related to the Landau quantization. The weak localization  gives NMR which is  often almost linear in a certain field range.   But such NMR
smoothly evolves from a sub-linear magnetic field dependence
at lower temperatures to super-linear field
dependence at higher temperatures, while we observe a perfectly linear NMR at low $B$ in a wide temperature range, Fig.(\ref{mrx}).
A strong NMR exists in
 the classical two-dimensional
electron gas due to freely circling electrons, which are not
taken into account by the Boltzmann approach. However, it is
parabolic  rather than linear  at low fields \cite{dyakonov}. The parabolic orbital NMR has been also predicted by the gauge theory in two-dimensional strongly-correlated doped Mott insulators \cite{wiegmann}.

On the other hand, if there is no time-reversal symmetry breaking,   the states with
the opposite direction of the orbital angular momentum, $m$ and
$-m$, are degenerate, so that the linear term in the MRh, Eq.(\ref{expansion}) cancels. But ions that carry a magnetic
moment break the time-reversal symmetry and split $m$ and $-m$
states. Such zero-field splitting (ZFS) gives preference to the
hopping via orbitals with a lower ionization energy (positive $m$)
providing the \emph{negative linear} MR \cite{aleomr}. In fact, we measured  ferromagnetic $M(B)$ hysteresis loop for \textbf{B} parallel to graphene planes in the electrically homogeneous graphite  (not shown here, see Ref. \cite{kopferro})
 demonstrating a metallic-like zero-field resistance in all  directions at temperatures below 50 K. MR of those samples can be  well fitted by $B^{4/3}$ law as in Fig.(\ref{fit}).

\begin{figure}
\begin{center}
\includegraphics[angle=-0,width=0.49\textwidth]{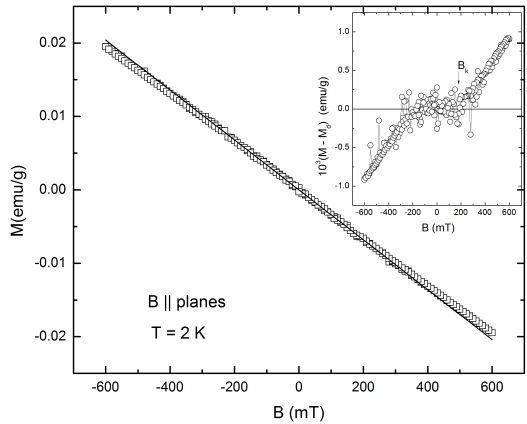}
\vskip -0.5mm \caption{  Diamagnetic magnetization of electrically inhomogeneous graphite  with NMR. Inset shows the  magnetization after subtracting the initial ($B < B_k$) linear diamagnetic contribution, $M_d(B)= \chi B$ with $\chi=-3.4 \times 10 ^{-5}$ emu/g mT.} \label{mag}
\end{center}
\end{figure}

However, we do not observe a global ferromagnetism in the bulk electrically inhomogeneous samples with the linear NMR, Fig.(\ref{mrx}). In a sharp contrast with the electrically homogeneous  samples the electrically inhomogeneous samples show a large diamagnetic response almost linear in $B$, Fig.(\ref{mag}).  Naturally, some  local paramagnetic centers responsible for the broken TRS could be  found also in the inhomogeneous graphite, with their magnetic response  overwhelmed  by the large diamagnetism.  However the observed large diamagnetism itself suggests a more intriguing mechanism of  TRS breaking  in our inhomogeneous samples, such as    superconducting clusters \cite{kopsp} with an unconventional (chiral)  order parameter \cite{don,chub}.  There is a kink in the field dependence of the diamagnetic magnetization of the inhomogeneous samples at $B_k \approx 0.2$ T, Fig.(\ref{mag}), resembling the behavior of  type-II superconductors in magnetic fields  exceeding the lower critical field  $B_{c1}$. Supporting this possibility the electrically inhomogeneous samples are becoming perfectly homogeneous insulators in sufficiently high magnetic fields, which could suppress the superconductivity, Fig.(\ref{tempdep}). Also in some   2D lattices there is the broken time-reversal
symmetry (i.e. some internal magnetic ordering) in the normal state with a spontaneous quantum Hall effect but without any net magnetic
flux  at high temperatures \cite{haldane}. At lower temperatures a pair of spontaneously
generated current loops in adjacent graphene layers, having odd-parity with respect to the two layers, could also break TRS \cite{varma}.

In conclusion, we presented  the highly-anisotropic  in-plane magnetoresistance in  graphite, which, in some current direction, is  negative and linear  in low fields  with the crossover to the positive MR at higher fields, while in the perpendicular current direction we observed the giant super-linear   positive MR.  The extraordinary MRs have been explained by the hopping magneto-conductance via non-zero angular momentum orbitals,  and by the  magneto-conductance of inhomogeneous media.  More generally our findings point to an inhomogeneous doping and a semiconducting gap in graphite (see also Ref.\cite{garcia}).
The linear orbital NMR is a unique signature of   the time-reversal symmetry breaking.  Combined with the large diamagnetism it points towards superconducting clusters with unconventional order parameter and normal state spontaneous currents in electrically inhomogeneous graphite samples.

This work has been supported by FAPESP, CNPq, CAPES, ROBOCON, INCT NAMITEC, the European Union Framework Programme 7
(NMP3-SL-2011-263104-HINTS), and
 by the UNICAMP visiting professorship programme.  We
thank Alexander Bratkovsky,  Alek Dediu,  Viktor Kabanov and Pavel Wiegmann for enlightening  discussions.

\end{document}